\documentclass[aps,prd,nofootinbib,twocolumn,superscriptaddress,preprintnumbers,balancelastpage,longbibliography,nobibnotes]{revtex4-1}
\usepackage{graphicx,epsfig,psfrag,bm,amssymb}
\usepackage{dcolumn}
\usepackage{bm}
\usepackage[dvipsnames]{xcolor}
\usepackage{braket}
\usepackage{mathrsfs,amsfonts,hepunits,color}
\usepackage{hyperref}
\usepackage{comment}
\usepackage{url}
\usepackage{soul}
\usepackage[]{natbib}
\usepackage[normalem]{ulem}

\newcommand{\be}{\begin{equation}}
\newcommand{\ee}{\end{equation}}
\newcommand{\bea}{\begin{eqnarray}}
\newcommand{\eea}{\end{eqnarray}}

\definecolor{linkcolor}{rgb}{0.0, 0.28, 0.67}
\hypersetup{colorlinks=true,
linkcolor=black,
citecolor=linkcolor,
urlcolor=linkcolor,
linktocpage=true
}

\newcommand{\E}{\mathbf{E}}
\newcommand{\B}{\mathbf{B}}
\newcommand{\D}{\mathbf{D}}
\newcommand{\h}{\mathbf{H}}

\newcommand{\Be}{\mathbf{B}_e}
\newcommand{\x}{\mathbf{x}}
\newcommand{\g}{g_{a\gamma\gamma}} 
\newcommand{\m}{m_a} 
\newcommand{\gr}{\Gamma_\rho}
\newcommand{\gp}{\Gamma}
\newcommand{\wLO}{\omega_{\rm LO}}
\newcommand{\wTO}{\omega_{\rm TO}}
\newcommand{\wpl}{\omega_{\rm p}}
\newcommand{\ez}{\epsilon_{0}}
\newcommand{\ei}{\epsilon_{\infty}}

\newcommand{\bepsilon}{\bm{\epsilon}}
\newcommand{\bmu}{\bm{\mu}}
\newcommand{\bsigma}{\bm{\sigma}}

\newcommand{\jeff}{\mathbf{J}_\text{eff}}
\newcommand{\rhoeff}{\rho_\text{eff}}
\newcommand{\rhof}{\rho_\text{f}}
\newcommand{\jf}{\mathbf{J}_\text{f}}

\newcommand{\w}{\omega}

\begin{document}

\preprint{KCL-PH-TH/2022-51}
\preprint{CP3-22-46}
\preprint{FERMILAB-PUB-22-736-T}
\title{Axion detection with phonon-polaritons revisited}

\author{David J. E. Marsh}
\affiliation{Department of Physics, King's College London, Strand, London, WC2R 2LS, United Kingdom}
\author{Jamie I.~McDonald}
\affiliation{Centre for Cosmology, Particle Physics and Phenomenology (CP3),
Université Catholique de Louvain,
Chemin du cyclotron 2,
Louvain-la-Neuve B-1348, Belgium}

\author{Alexander J. Millar}
\affiliation{The Oskar Klein Centre, Department of Physics, Stockholm University, AlbaNova, SE-10691 Stockholm, Sweden}
\affiliation{Nordita, KTH Royal Institute of Technology and
Stockholm
  University, Roslagstullsbacken 23, 10691 Stockholm, Sweden}
  \affiliation{Fermi National Accelerator Laboratory, Batavia, Illinois 60510, USA}
\author{Jan Sch\"{u}tte-Engel}
\affiliation{Department of Physics, University of Illinois at Urbana-Champaign, Urbana, IL 61801, USA}
\affiliation{Illinois Center for Advanced Studies of the Universe, University of Illinois at Urbana-Champaign, Urbana, IL 61801, USA}
\date\today

\begin{abstract}
In the presence of a background magnetic field, axion dark matter induces an electric field and can thus excite phonon-polaritons in suitable materials. We revisit the calculation of the axion-photon conversion power output from such materials, accounting for finite volume effects, and material losses. Our calculation shows how phonon-polaritons can be converted to propagating photons at the material boundary, offering a route to detecting the signal. Using the dielectric functions of GaAs, Al$_2$O$_3$, and SiO$_2$, a fit to our loss model leads to a signal of lower magnitude than previous calculations. We demonstrate how knowledge of resonances in the dielectric function can directly be used to calculate the sensitivity of any material to axion dark matter. We argue that a combination of low losses encountered at $\mathcal{O}(1)$ K temperatures and near future improvements in detector dark count allow one to probe the QCD axion in the mass range $m_a\approx 100$ meV. This provides further impetus to examine novel materials and further develop detectors in the THz regime. We also discuss possible tuning methods to scan the axion mass.
\end{abstract}

\maketitle

\section{Introduction}

The search for axion dark matter (DM) is gathering pace around the world~\cite{Chadha-Day:2021szb,Semertzidis:2021rxs}. A consistently troubling part of parameter space lies at high frequency, $\nu\gtrsim 100\text{ GHz}$ corresponding to axion masses $m_a\gtrsim 1\text{ meV}$, where the tried and tested microwave cavity haloscopes~\cite{Sikivie:1983ip,ADMX:2021nhd,McAllister:2017lkb,HAYSTAC:2020kwv,Semertzidis:2019gkj,TASEH:2022vvu} cease to be a viable technology. There are three distinct challenges at high frequency: building (wide band) detectors, building (large volume) resonators, and resonator tunability. The dish antenna approach~\cite{Horns:2012jf} taken by BRASS~\cite{Horns:2012jf} and BREAD~\cite{BREAD:2021tpx} sidesteps the second and third problems and collects an unamplified signal across a wide frequency range and long integration time. However, a range of detectors are still needed to cover the putative wide band, and resonators can still play a complementary role.

A dielectric haloscope~\cite{Caldwell:2016dcw} (related to open resonators~\cite{Cervantes:2022epl}) enhances the signal relative to a dish antenna by a ``boost factor'', $\beta$, such that the power output (more details are given below) is:
\be
P = \beta(\omega)^2 P_{\rm dish}\, .
\ee
The boost factor concept is sufficiently general that it can be applied to any would-be axion detection technology, and we use this approach as our metric in the following. The boost factor can have a complex dependence on frequency, $\omega$, and may be resonant, or achieve more broadband enhancement, depending on the layout of the dielectric media~\cite{Millar:2016cjp}. A related concept is the wire meta-material plasma haloscope~\cite{Lawson:2019brd}, which achieves optimal resonant signal enhancement independent of volume. It is believed that dielectric and plasma haloscopes can be mechanically tuned for frequencies up to 100 GHz, though prototypes are still in development~\cite{Egge:2020hyo,Wooten:2022vpj,Balafendiev:2022wua}. 

Dielectric haloscopes have been proposed~\cite{Baryakhtar:2018doz} and prototypes implemented~\cite{Chiles:2021gxk,Manenti:2021whp} in the optical frequency range, but they still face problems at intermediate, THz, frequencies. One route to THz resonant axion technology is the use of axion-polaritons~\cite{Marsh:2018dlj,Schutte-Engel:2021bqm,Chigusa:2021mci}. Axion-polaritons are formed by mass-mixing between magnetic axion-quasiparticles and the (free) electric field inside topological insulators, and can in principle be tuned via an applied magnetic field. A material realising such quasiparticles has unfortunately yet to be conclusively demonstrated in transmission measurements as required by Ref.~\cite{Schutte-Engel:2021bqm}. Mn$_2$Bi$_2$Te$_5$, however, remains a promising candidate with significant recent progress being~\cite{2020ChPhL..37g7304Z,PhysRevB.104.054421}. As yet unknown losses in these materials represent the largest uncertainty about their viability in DM searches.

The idea of axion DM detection typified by the axion quasiparticle proposal is essentially captured, however, not by the exotic axion quasiparticles themselves, but by the polariton dispersion relation: mass mixing between a particle-like excitation and the electromagnetic field, which can then be resonantly driven by the axion DM source. The polariton must occur in a low temperature, low loss system. One suitable polariton available in condensed matter systems is the phonon-polariton~\cite{Mills_1974}. Axion detection in phonon-polariton materials was proposed in Ref.~\cite{Mitridate:2020kly}. Phonon-polariton materials are difficult to tune, although we suggest some possibilities. Resonant materials could also be substituted into dish-antenna type setups to achieve signal enhancement in different frequency ranges using a range of materials. 

In Ref.~\cite{Mitridate:2020kly}, the axion DM signal in phonon-polariton materials was computed from the scattering cross-section, and volume scaled, giving event rates in (kg yr)$^{-1}$ and assuming a background free measurement. The problem with this approach when applied to axion searches is the ability to read out the resulting signal, which would require single phonon detection in the bulk of the material, or measuring total heat deposition, or similar. Instead, we focus on detecting an emitted photon, which can use similar detection technology and setups to existing methods. Such an approach can also enhance the resonance by making use of resonant boundary conditions. We show, furthermore, that accounting for the decay length of phonon-polaritons limits the maximum useful material volume.

We take Ref.~\cite{Mitridate:2020kly} as a starting point and consider the phonon-polariton materials AL$_2$O$_3$, GaAs and SiO$_2$ for which properties are collated and tabulated by the authors of Ref.~\cite{Knapen:2021bwg}. Our treatment differs from Ref.~\cite{Mitridate:2020kly} in that we compute the axion-photon conversion power output in a classical field theory calculation in exact analogy to the dielectric haloscope, plasma haloscope, and axion-quasiparticle haloscope calculations of Refs.~\cite{Millar:2016cjp,Lawson:2019brd,Schutte-Engel:2021bqm}. By considering the material in a finite volume, and properly accounting for boundary conditions, we are able to show how axion-induced phonon-polaritons convert to propagating photons at the boundary, offering a more standard means of detection with e.g. photon counters, of which we consider various possibilities. We furthermore include losses in phonon-polariton materials according to measurements and simulations in \cite{Knapen:2021bwg} (detailed references are given in table \ref{tab:losses}).
Finally, we discuss possibilities to enhance the signal if losses reduce to a minimum set by impurity density and phonon decays at low temperature, and discuss the possibility to tune the resonance.

This paper is organized as follows: in section~\ref{sec:Axion_electrodynamics} we describe how axion DM can resonantly excite phonon-polaritons for both a thin slab and large volume of material. The emitted electromagnetic fields that are emitted from the material can be resonantly enhanced, while the amount of enhancement depends on the losses in the respective materials. In section~\ref{sec:sensitivity} we show the sensitivity of the three candidate materials to axion dark matter with respect to the axion photon coupling. We conclude in section~\ref{sec:conclusion}.

\section{Axion Electrodynamics in Phonon-Polariton Media}\label{sec:Axion_electrodynamics}
To calculate the power produced by axions converting to phonon-polaritons we can simply solve the classical axion-Maxwell equations. Even at $100\,$meV the occupation number of axions is very high, ${\cal O}(10^{10})$, lending itself to a classical field description. Such a calculation can easily be compared with a quantum mechanical one (such as in Ref.~\cite{Mitridate:2020kly}) by noting that the classical result simply gives the expectation value of the quantum mechanical rate~\cite{Raffelt:1987im,Ioannisian:2017srr}.

\subsection{Axion Electrodynamics in Media}

The axion-Maxwell equations in a material are of the form:
\bea
\nabla\cdot \D&=&\rhof+\rhoeff,\label{eq:Gauss_E_law}\\
\nabla\times \h - \partial_t\D&=&\jf+\jeff,\label{eq:Ampere_law}\\
\nabla\cdot \B&=&0,\label{eq:Gauss_B_law}\\
\nabla\times\E+\partial_t \B&=&0,\label{eq:Faraday_law}
\eea

where $\h$ is the magnetic field, $\B$ the magnetic flux density, $\E$ the electric field, $\D$ the electric displacement field, $\rhof$ the free charge density, $\jf$ the free current density. $\rhoeff$ and $\jeff$ are the effective charge and current densities that are induced by the axion-photon mixing. We assume that the dark matter axions are non-relativistic and have mass $\m$. They are hence described by a field $a(t)$. Furthermore we assume a constant external $B$-field $\Be$. For phonon materials the free charge density in the material vanishes. Altogether we can simply write the effective axion current density as $\jeff(t)=\g\Be \partial_t a$, where $\g$ is the axion-photon coupling. The effective charge density $\rhoeff$ vanishes.

We Fourier transform eqs.~\eqref{eq:Gauss_E_law}-\eqref{eq:Faraday_law} in the time domain and use the relations:
\bea
\D(\x,\w)&=&\bepsilon(\omega)\E(\x,\w),\\
\textbf{H}(\x,\w)&=&\bmu^{-1}(\omega)\B(\x,\w),\\
\jf(\x,\w)&=&\bsigma(\omega)\E(\x,\w),
\eea
where $\bepsilon$ is the permittivity tensor, $\bsigma$ the conductivity and $\bmu$ the permeability tensor, which we will set to one in the following, since we consider phonon materials and not magnetic materials. We want to point out that our treatment could also be extended to include a non-trivial permeability tensor. We finally end up with an equation for the electric field:
\bea
-\nabla\times\nabla\times\E(\x,\w)+\w^2\left(\bepsilon(\w)+i\frac{\bsigma(\w)}{\w}\right)\E(\x,\w)\nonumber\\=-\w^2 \g\Be a(\w).\label{eq:wave_eq_E}
\eea
We assume in the following that the material is aligned with the external $B$-field such that the off diagonal elements of $\bepsilon$ and $\bsigma$ are small compared to the diagonal elements: $\bepsilon(\omega)={\rm Diag}(\epsilon_{xx},\epsilon_{yy},\epsilon_{zz})$, $\bsigma(\omega)={\rm Diag}(\sigma_{xx},\sigma_{yy},\sigma_{zz})$.
Note that off diagonal $\bepsilon$ and $\bsigma$ components can also excite polarizations that are not parallel to $\Be$. However these effects are expected to be small compared to the effects that we describe here when properly aligned. 
A detailed experiment design should include a full dielectric tensor treatment. 

Assuming that the external $B$-field is polarized in $y$-direction and the problem is one dimensional, i.e. all fields depend only on the $z$-coordinate, the equation for the $y$-component of the $E$-field is:
\be
\partial_z^2 E(z,\w)+\omega^2n^2(\w) E(z,\w)=-\w^2\g B_ea(\w),
\label{eq:E_pde_1D}
\ee
where we have dropped the subscripts for the $E$-field as well as for the $\epsilon$ and $\sigma$ components and we have defined:
\be
n^2(\w):=\epsilon(\w)+i\frac{\sigma(\w)}{\w}.
\ee
In deriving eq.~\eqref{eq:E_pde_1D} we have used that $\nabla \cdot \E=0$, which is true if we set the $z$-component of the $E$-field to zero and assume that the problem is one-dimensional in the $z$-direction.
The last assumption is justified because the material size in the $z$-direction is much smaller than in the $x$ and $y$-directions. Note that with our assumptions eq.~\eqref{eq:Gauss_E_law} is also fulfilled since there is typically no free charge density $\rhof$ in phonon materials.

\begin{figure}
    \centering
    \includegraphics[width=0.3\textwidth]{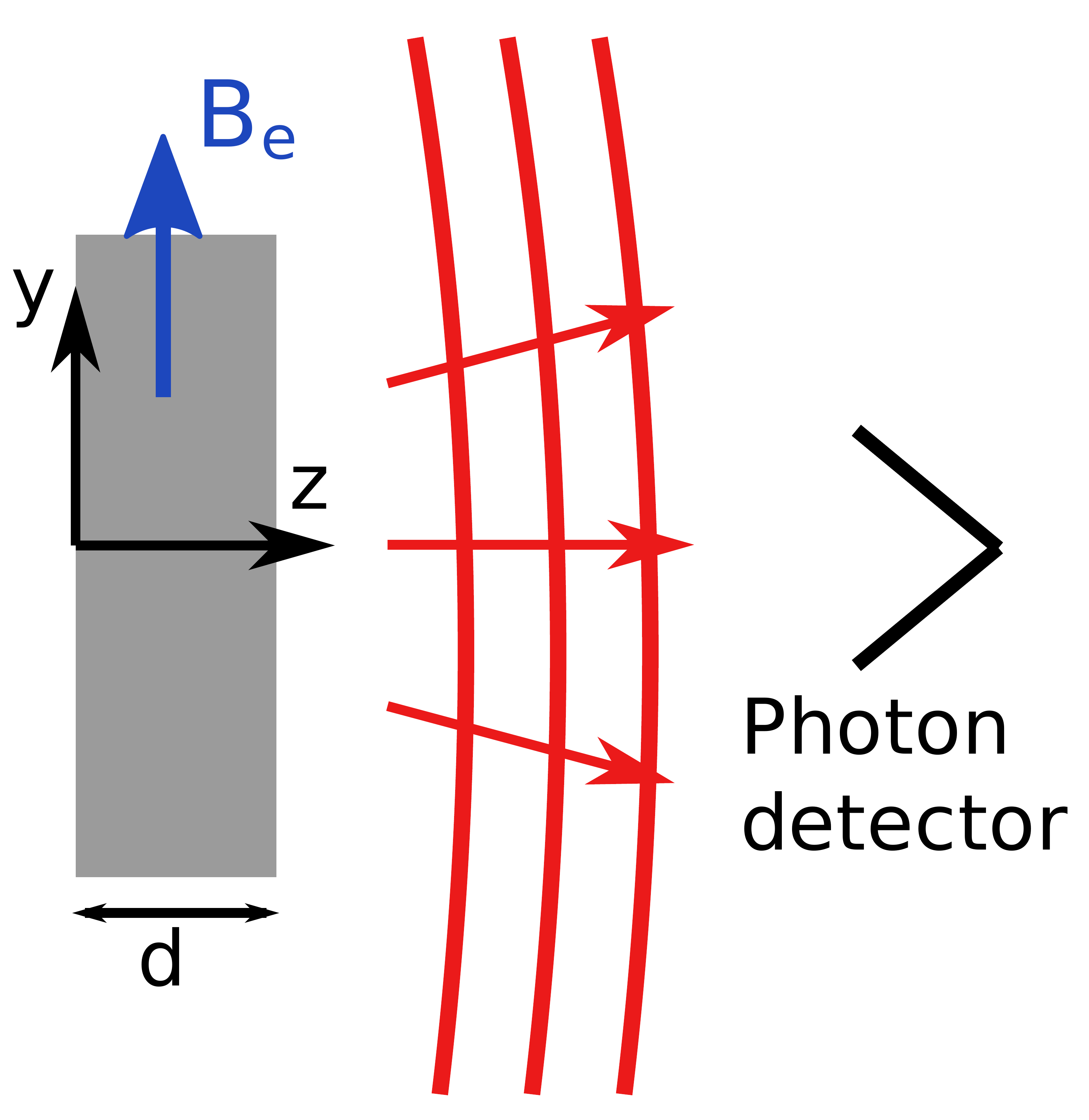}
    \caption{We show a phonon material of thickness $d$ that emits electromagnetic radiation (in red) that is detected by a single photon detector. The external $B$-field is aligned in the $y$-direction.}
    \label{fig:setup}
\end{figure}

The solution to the eq.~\eqref{eq:E_pde_1D} is:
\be
E(z,\w)=-\frac{a(\w)B_e \g}{n^2(\w)}+C_+ e^{ikz}+C_- e^{-ikz},
\ee
with $k=\w \,n(\w) $ and $C_\pm$ are constants that can be determined with appropriate interface conditions.
We now consider a material slab of thickness $d$ that is surrounded by vacuum, cf. Fig.~\ref{fig:setup}. 

The emitted electromagnetic field from the material slab of thickness $d$ can be obtained by matching the $E$ and $B$-field in vacuum and in the material. Similar calculations as in Ref.~\cite{Millar:2016cjp,Schutte-Engel:2021bqm} yield an outgoing field with amplitude
\begin{eqnarray}
E_{\rm out}=E_0\frac{(n^2(\omega)-1)\sin(\Delta/2)}{n^2(\omega)\sin(\Delta/2)+in(\omega)\cos(\Delta/2)},
\label{eq:E_out}
\end{eqnarray}
where $E_0=\g B_e a(\w)$ and $\Delta=d  \w n(\w) $.
The emitted electromagnetic power from the material is
$P_{\rm sig}=\frac{1}{2}\left|E_0\right|^2\beta^2A$,
where $A$ is the surface of the material in the $xy$-plane and $\beta$ is the boost factor:
\be
\beta=\left|\frac{E_{\rm out}}{E_0}\right|.
\ee
The boost factor $\beta$ can be significantly enhanced if there is a resonance in the material.
One of the main points of the present work is that the resonances are all encoded in the form of the dielectric function, i.e. refractive index. We will show that if the refractive index $n(\w)$ is known then one can directly specify the sensitivity reach of the specific material to the axion-photon coupling $\g$.

\subsection{Classical phonon-polariton dispersion relation}\label{sec:dispersion}

In a first scenario we assume a classical phonon-polariton dispersion relation with one resonance~\cite{Mills_1974}:
\begin{align}
       n^2(\w)&=\epsilon(\w)+i\frac{\sigma}{\w} \nonumber \\
       &=\ei\left(1+\frac{\wpl^2/\ei}{\w_{\rm TO}^2-\w^2-i\w\gp}+i\frac{\gr}{\omega}\right)\, .
       \label{eq:dispersion_relation_simple}
\end{align}
Phonon-polaritons couple an electronic lattice degree of freedom (phonons) to the electric field. Conductivity leads to electric field losses, while phonons have thermal scattering losses, decays, and losses related to lattice impurities. Thus, the resonance is damped by two loss parameters $\gr$ and $\gp$ which describe photon and phonon losses respectively. $\ei$ is the permittivity for $\w\gg \wTO$, $\wpl$ is the effective unscreened plasma frequency and $\wTO$ is the transverse optical phonon frequency. The photon loss parameter can be expressed through the conductivity and $\ei$ as $\gr=\sigma/\ei$.

In the following we discuss the appearance of a resonant enhancement of the boost factor. The discussed dielectric function is typical for a phonon-polariton material. Therefore the discussion here is very similar to the discussion of axion-polariton materials, cf. Ref.~\cite{Schutte-Engel:2021bqm}. We can therefore overtake large parts of the calculations that have been done in Ref.~\cite{Schutte-Engel:2021bqm}. From Eq.~\eqref{eq:E_out} and Eq.~\eqref{eq:dispersion_relation_simple} it becomes clear that the resonance condition is:
\bea
\Delta = \Delta_j=n(\w_j) \, \w_j d= (2j+1) \, \pi \, , \quad j\in\mathbb{N}_0.
\label{eq:resonance_condition}
\eea

The resonance frequencies are called $\w_j$ and can be expressed as follows:
\bea
	\w_j^2&=\frac{\wLO^2}{2} +  
	 \frac{\Delta^2_j}{2d^2\ei}  + \sqrt{ \frac{\wpl^2 \Delta _j^2}{d^2} + \left(\frac{\wLO^2}{2}-\frac{\Delta_j^2}{2 d^2 \ei}  \right)^2} \, ,\nonumber\\
	 &=\wLO^2+\delta\omega_j^2+\mathcal{O}\left(\frac{4\Delta_j^2\wpl^2}{d^2\wLO^4}\right) \, ,
\label{eq_AQ_DA_P_mixing_res_frequencies}
\eea
where we have defined the longitudinal optical frequency $\wLO^2=\wTO^2+\wpl^2/\ei$ and 
\bea
\delta\w_j^2\equiv\frac{\Delta_j^2\wpl^2}{d^2\wLO^2} \, .
\eea
This corresponds to the rest mass of the phonon-polariton and is distinct from $\omega_{\rm TO}$, which could cause a potential signal to be missed if the detector is only looking at $\omega_{\rm TO}$ or the data analysis assumes that $\omega_{\rm LO}=\omega_{\rm TO}$ (for example, for GaAs $\omega_{\rm LO}=1.09\omega_{\rm TO}$).  In Ref.~\cite{Mitridate:2020kly} this effect was prescribed to the splitting between transverse and longitudinal phonons, with the polaritons being degenerate with the longitudinal phonons. 


We can expand the emitted electromagnetic field around the resonance frequencies:
\begin{eqnarray}
	\frac{E_{\rm out}}{E_0}=-\frac{iA_j}{i\gamma_j\w_j+(\w^2-\w_j^2)} \, ,
\label{eq:DA_AQ_P_mixing_LayerBoost_res_simp}
\end{eqnarray}
with
\bea
\gamma_j&=&\frac{4\wpl^2\Delta_j^2}{\wLO^4d^3}+\left(\gp+\frac{\wpl^2}{\ei\wLO^2}\gr\right)\, \label{eq:gammaloss} \, ,\label{eq:TwoInterfaces_FWHM} \\
A_j&=&\frac{4\wpl^2}{\w_j d}\approx \frac{4\wpl^2}{\wLO d} \, ,
\eea
where we used that in a resonant case $\w_j$ is close to $\wLO$. We can see that the width of the resonance decreases rapidly as a function of $d$ until losses take over. To estimate the sensitivity to errors in the thickness of the material we can use the analytic formula derived in Ref.~\cite{Millar:2016cjp}. For a 10\% variation in $\beta^2$ the Gaussian standard deviation in thickness $\sigma$ should be
\begin{equation}
\sigma\lesssim 20\,{\rm nm}\left (\frac{10^2}{\beta}\right)^{1/2}\left(\frac{100\,{\rm meV}}{m_a}\right).
\end{equation}
Thus for lower loss, i.e. more resonant systems, the surfaces must be manufactured to a higher degree of smoothness. 

While such numbers should be achievable, it is also possible to avoid the issue. In this calculation we assumed that the material was relatively thin, so that a standing wave is formed. However, if the thickness is much larger than the decay length in the medium ($1/(\Gamma+\omega_p^2\Gamma_\rho/\epsilon_\infty\omega_{\rm LO}^2)$ then it will behave as a half infinite slab. One can see from Ref.~\cite{Millar:2016cjp} that such a system would have 
\begin{equation}
\frac{E_{\rm out}}{E_0}=1-\frac{1}{\sqrt{i\left(\frac{\Gamma_\rho}{\omega_{\rm LO}}+\epsilon_\infty^2\frac{\omega_{\rm LO}\Gamma}{\omega_p^2} \right)}}\,.
\end{equation}

The resonance frequencies $\w_j$ can in principle be tuned by applying a pressure and/or strain to the materials. For example in Refs.~\cite{PhysRevB.41.10104,PhysRevB.54.2480} it was shown that $\epsilon_\infty$ can be changed up to $10\%$ when a maximal pressure of $1\,$GPa is applied. Higher pressures are not considered since they can lead to significant changes in the materials, i.e. phase transitions. If one assumes that the other parameters that enter in the expression for the resonance frequency stay constant, then the resonance frequency can change up to approximately $5\%$. This demonstrates that in principle the application of pressure/strain can change the resonance frequency. In reality also other parameters than $\epsilon_\infty$ may change when a pressure/strain is applied and one might get larger modifications of the resonance frequency.
A change in the resonance frequency allows some tuning in the axion search to scan different masses. 

Pressure tuning might, however, not be practical for two reasons. Firstly, high pressures are needed to get a significant change in the material parameters, which may present an engineering challenge. Secondly, the pressure might have to be applied to the material in such a way that it hinders the detection of the outgoing photons. We propose that strain is also investigated as a tuning mechanism, and expect that similar 10\% tuning might be possible.

\subsection{Losses in phonon-polariton materials}\label{sec:phonon_Losses}

We now estimate the losses and resonances in candidate materials. We consider the following materials, following Ref.~\cite{Mitridate:2020kly}: GaAs (gallium arsenide), Al$_2$O$_3$ (\emph{corunudrum} aluminium oxide), SiO$_2$ (\emph{alpha quartz}, silicon dioxide). We would like to infer the values of loss parameters based on fitting the expression Eq.~\eqref{eq:dispersion_relation_simple} to experimental data. In Ref.~\cite{Knapen:2021bwg} (and references therein quoted in Table \ref{tab:losses}) the fitting was performed neglecting the electrical conductivity in Eq.~\eqref{eq:dispersion_relation_simple}, i.e. setting $\Gamma_\rho =0$. Fitting was therefore based on a sum over simple phonon resonances of the form $\epsilon(\omega)$. These fits were in good agreement with data assuming no electrical conductivity - see Fig.~1 of Ref.~\cite{Gervais_1974}. This is consistent with the losses from photons being sub-dominant compared to those from phonons ($\Gamma \gg \Gamma_\rho$). Conductivity is the inverse of the resistivity $\rho$ so that assuming $\epsilon_\infty =\mathcal{O}(1)$ we have $\Gamma_\rho \sim 1/\rho = 0.6\text{ meV}\,[\rho/(\Omega\text{cm})]^{-1}$. A high quality factor $Q=\omega/\Gamma\gg 1$ in the phonon-polariton frequency range $\omega\sim 50\text{ meV}$ requires large resistance $\rho\gg 1 \,\Omega\text{cm}$. From the quality of the fits we see that this is indeed  true at the temperatures of the measurements performed. We will assume this condition continues to hold at lower temperatures than those in Table \ref{tab:losses} so that photon losses can also be neglected at sub-K temperatures encountered in axion experiments. Ultimately, what is needed are new measurements at cryogenic temperatures.

Let us then discuss the remaining losses associated to phonons. As lattice degrees of freedom, in analogy to magnons~\cite{PhysRevLett.111.017204}, we consider three loss channels for phonons: scattering, impurities, and decays. General arguments based on the Boltzmann equation (see Ref.~\cite{Schutte-Engel:2021bqm} for a brief review) suggest that scattering losses in general decrease with temperature with some strong power law, and are exponentially suppressed for $T<\omega_{\rm phonon}$, where $\omega_{\rm phonon}$ is the rest-frame phonon frequency. At very low temperatures for stable resonances, impurities dominate the lifetimes of lattice-coupled degrees of freedom, and the zero-temperature width can be estimated as:
\be
\frac{\Gamma_{\rm imp.}}{\omega_{\rm LO}}\approx \frac{\delta L}{L}\, ,
\label{eqn:gamma0}
\ee
where $\delta L$ is the lattice spacing, and $L$ is the typical crystal grain size, which we take to be around 1$\mu$m (for sapphire, see e.g. Ref.~\cite{1999PhLA..261....5U}). Estimates for this contribution to the width are given in Table \ref{tab:losses}. For Al$_2$O$_3$ and SiO$_2$, which have hexagonal lattices, we use the larger $b$ lattice constant to give the largest, and thus most conservative, width due to impurities.

The phonons under consideration are ``optical phonons'', so-called because they couple to the electromagnetic field, as required in the present case to form the phonon-polariton. Optical phonons have the required property of a dispersion relation the same as a massive particle, while acoustic phonons behave as massless particles. Optical phonons couple to the acoustic phonons via the interatomic potential, and thus can decay into them. The optical phonon width due to such a decay was computed for GaAs in Ref.~\cite{1994JAP....76.3905B}, and generalised in Ref.~\cite{PhysRevLett.88.215502}. The LO phonon lifetime at zero temperature in GaAs is computed to be $\tau_{\rm dec.}=10^{-11}\text{ s}$~\cite{1994JAP....76.3905B}, corresponding to a width $\Gamma_{dec.}=1/\tau_{\rm dec.}=0.066\text{ meV}$, and thus $\Gamma_{\rm dec.}/\omega_{\rm LO}\approx 1.8\times10^{-3}$, around a factor of three larger than the contribution from impurities.

In the general case, the matrix element for optical to acoustic phonon decay is the Fourier transform of the third spatial derivative of the interatomic potential~\cite{PhysRevLett.88.215502}. In order to estimate the decay width for Al$_2$O$_3$ and SiO$_2$ from the GaAs calculation~\cite{1994JAP....76.3905B} we assume that the derivatives scale like the lattice constants, i.e. $\partial V/\partial x_i\sim (\delta L_i)^{-1}$. GaAs has a cubic lattice while our other materials have a hexagonal lattice. For our rough estimate of the scaling, we use the smaller $a$ lattice constant twice in Eq.~\eqref{eqn:decay_rescale} to get the larger, more conservative, estimation of the width. We further assume that the width scales with the LO resonance frequency. Thus for a material $x$ our estimate for the width is:
\begin{equation}
    \frac{\Gamma_{{\rm dec},x}}{\omega_{\rm LO}} = \frac{\Gamma_{{\rm dec.,GaAs}}}{\omega_{\rm LO}}\left[\frac{\delta L_{a,{\rm GaAs}}}{\delta L_{a,x}}\frac{\delta L_{a,{\rm GaAs}}}{\delta L_{a,x}}\frac{\delta L_{a,{\rm GaAs}}}{\delta L_{b,x}}\right]^2\, .
    \label{eqn:decay_rescale}
\end{equation}
Estimates for this contribution to the width are given in Table \ref{tab:losses}.

Our fits for the widths from existing measurements are taken at $T= \mathcal{O}(10-100)\text{ K}$. If at these temperatures the measured phonon linewidth is larger than the value estimated from crystal impurities or phonon decays, i.e. $\Gamma>\Gamma_{\rm imp.,dec.}$, we expect that going to lower temperatures will lead to narrower linewidths (lower losses). Modern axion search experiments typically operate at sub-K temperatures, and we therefore take the larger of $\Gamma_{\rm imp.,dec.}$ as our most optimistic phonon-polariton linewidth.

\begin{table*}
	\center
\bgroup
\def\arraystretch{1.7}
\begin{tabular}{ l l l l }
\hline Material & $\Gamma$ [meV]   &   $\Gamma_{\rm imp.}/\omega_{\rm LO}$  &  $\Gamma_{\rm dec.}/\omega_{\rm LO}$ \\ 
\hline
\hline  Al$_2$O$_3$ (sapphire, 77 K ~\cite{Gervais_1974})  \qquad &  0.19, 0.39    & ${ 1.3\times 10^{-3}}$ & $6.9\times10^{-4}$  \\
GaAs (10 K ~\cite{LawlerShirely,Palik}) & 0.31     & $5.7\times 10^{-4}$ & $1.8\times 10^{-3}$ \\
  SiO$_2$ ($\alpha$-quartz, 300K \cite{GervaisTiO2}) & 0.59, 0.63, 0.72, 0.82 \quad & $4.9\times 10^{-4}$ & $3.7 \times10^{-3}$ \\
\hline 
\end{tabular}
\egroup
\caption{Phonon polariton loss estimates. The first column gives the measured width from fitting Eq.~\eqref{eq:dispersion_relation_simple}. Data was taken from tables collated by the authors of \cite{Knapen:2021bwg}. We also give the explicit literature references for each material and the temperatures at which the data was produced either numerically or experimentally. We estimate the impurity contribution to the width, $\Gamma_{\rm imp.}/\omega_{\rm LO}$, using Eq.~\eqref{eqn:gamma0} assuming $L=1\,\mu\text{m}$. We estimate the decay contribution, $\Gamma_{\rm dec.}/\omega_{\rm LO}$, from Ref.~\cite{1994JAP....76.3905B} for GaAs, rescaled to the other materials using Eq.~\eqref{eqn:decay_rescale}. Lattice constants used for estimations are taken from MatWeb~\cite{MatWeb}.}
\label{tab:losses}
\end{table*}

\subsection{Volumetric detectors}

In this section we discuss a dielectric function that has been demonstrated to describe a particular material. In Ref.~\cite{doi:10.1063/1.363818} the dielectric function for gallium arsenide (GaAs) has been determined by fitting transmission measurements to a model for the dielectric function:
\begin{align}
n^2(\w)=& \epsilon(\w) \nonumber \\ =&\frac{(\epsilon_0-\epsilon_\infty)\,\wTO^2}{\wTO^2-\w^2+i\w\Gamma_{\rm ph}}
+\frac{(\epsilon_\infty-\epsilon_{\rm uv})\,\w_{\rm vis}^2}{\w_{\rm vis}^2-\w^2+i\w\Gamma_{\rm vis}}+\epsilon_{\rm uv},
\end{align}
where we have set $\sigma=0$ to simplify the discussion. 

In Ref.~\cite{Mitridate:2020kly} the system considered was taken to be very large relative to the decay length in the material, leading to an overall volumetric effect. However, as the calculation was performed from first principles it is generally intuitive to fit the results therein to characterizations of the material. The dielectric constant was taken to screen by a factor corresponding to $\epsilon_{\infty}$, where any frequency dependent behavior was hidden in a complicated rate formula. 

However, as the dark matter interacts with the photon, rather than the medium directly, the medium response of the photon is essentially all the dark matter sees. Thus by a simple frequency dependent dielectric constant one can capture all relevant behavior in an intuitive way that is simple to verify experimentally. To show this, we consider a large sample, where the signal power would be somehow readout from the bulk, as considered in Ref.~\cite{Mitridate:2020kly}.  

As the axion induced $E$-field is proportional to $1/\epsilon$, the resonance will occur when ${\rm Re}(\epsilon)=0$. As we are considering only the lowest pole, we can take $\omega\ll \omega_{\rm vis}$ to get
\begin{equation}
    \epsilon(\w)=\frac{(\epsilon_0-\epsilon_\infty)\,\wTO^2}{\wTO^2-\w^2+i\w\Gamma_{\rm ph}}
+\epsilon_\infty.
    \label{eq:n_GaAs_simplified}
\end{equation}
This formalism agrees with Eq.~\eqref{eq:dispersion_relation_simple} under the assumption that $(\ez-\ei)\wTO^2=\wpl^2$. 
The resonance condition is satisfied for a frequency $\omega_{\rm LO}$
\begin{equation}
    \omega_{\rm LO}=\sqrt{\frac{\epsilon_0}{\epsilon_\infty}}\omega_{\rm TO}.\label{eq:LST}
\end{equation}
Equation~\eqref{eq:LST} is referred to as the Lyddane–Sachs–Teller relation~\cite{PhysRev.59.673}.

On resonance the $E$-field in an infinite medium would be given by
\begin{equation}
    E=-i\frac{\epsilon_\infty-\epsilon_0}{\epsilon_\infty\epsilon_0}\frac{\omega_{\rm LO}}{\Gamma_{\rm ph}}E_0.
\end{equation}
We can calculate the power dissipated by using~\cite{landau2013electrodynamics}
\begin{equation}
P=\frac{\omega}{2}\epsilon''\int|E|^2dV,
\end{equation}
which, after seeing that $Q=\omega/\Gamma_{\rm ph}$, can be rearranged to get
\begin{equation}
    P=\g^2B_e^2V\rho_{\rm DM}\frac{Q}{\omega}\frac{\epsilon_\infty-\epsilon_0}{\epsilon_\infty\epsilon_0}.
\end{equation}
Note that this considers only the power dissipated in the medium, and neglects any coupling to an antenna or detector. For reference, the same expression for a plasma haloscope (neglecting boundary conditions) is given by $P=\g^2B_e^2V\rho_{\rm DM}Q/\omega$~\cite{Lawson:2019brd,Caputo:2020quz}. Thus we can see that while the width of the resonance in both cases is given by $Q$, the power dissipated in phonons is suppressed relative to the ideal case of an unscreened plasma by a factor $\frac{\epsilon_\infty-\epsilon_0}{\epsilon_\infty\epsilon_0}$. For GaAs this factor is 0.015, i.e., a two order of magnitude suppression, rather than the one order expected from $\epsilon_\infty$. However, as the analysis of Ref.~\cite{Mitridate:2020kly} treated the high and low frequency medium separately, rather than as part of a consistent dielectric function they only use an explicit screening factor of $1/\epsilon_{\infty}$. While both writings are correct, such a division is somewhat artificial.   

This difference from the naive screening term, which one would see in, for example a plasma with a non-unity dielectric constant at high frequencies is due to the dielectric function interpolating between two positive values, $\epsilon_0$ and $\epsilon_\infty$, rather than a transition between a positive dielectric constant and an infinite imaginary one (i.e., transitioning between a dielectric and a metal). This interpolation naturally leads to values closer to $|\epsilon|=0$. In the language of Ref.~\cite{Mitridate:2020kly} the screening has two factors, one coming from the static electron background and one coming from the phonon-photon interaction. 

However, a large volume detector of this nature raises the issue of how to extract a signal. Instrumenting the interior of the material would be very difficult as it would require inserting many THz antennas into the sample. This is because the losses give a corresponding decay length to the phonon-polariton ($1/\Gamma_{\rm ph}$), so phonon-polaritons would decay in the interior of the device. 

Thus, either only the power emitted from the surfaces can be read, in which case one can get more power by optimising a thin disk of material as described above, or some kind of calorimeter is needed. The lowest threshold current calorimeters are trying to reach ${\cal O}(100)$\,meV\cite{cite-key,Fink:2020noh}, making such a detection system difficult, especially for lower masses and higher target volumes. Thus which method proves more practical is highly dependent on available detection technology. 

\section{Experimental Sensitivity}\label{sec:sensitivity}
 
\begin{figure*}
    \centering
        \includegraphics[width=0.95\textwidth]{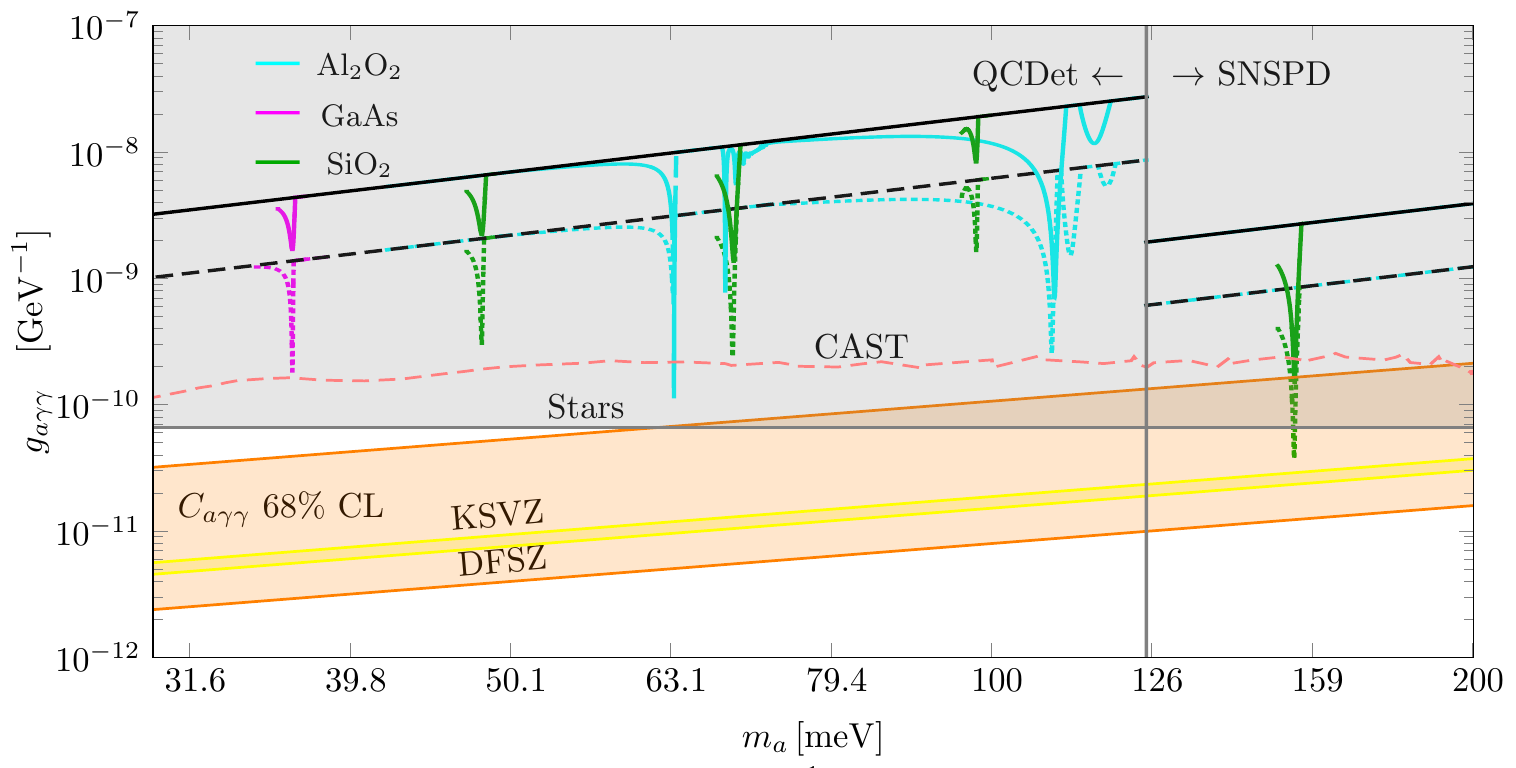}
    \caption{
    \textbf{Sensitivity.} 
     Expected 95\% exclusion limit on the axion photon coupling from phonon resonances for some materials considered in \cite{Mitridate:2020kly,Knapen:2021bwg}. For comparison we show the current limit set by stellar cooling \cite{Balazs:2022tjl,Dolan:2022kul} (gray block), and the CAST experiment \cite{CAST:2017uph} (dashed pink). We also show projections for a dish antenna (black) of equivalent area. The QCD model band defined by Ref.~\cite{Plakkot:2021xyx} is shown in orange. Solid and dashed lines respectively indicate conservative and improved scenarios detailed in the main text. 
    }
    \label{fig:Sensitivity}
\end{figure*}

If $n_s$ and $n_d$ are the signal and background events, then the significance is $S=2\left(\sqrt{n_s+n_d}-\sqrt{n_d}\right)$~\cite{doi:10.1142/S0217732398003442,Bityukov:2000tt,Arias:2010bh}. The $95\%$ exclusion limit is obtained by requiring $S=2$, i.e. $\lambda_s<\frac{1}{\tau}+2\frac{\lambda_d}{\tau}$, where $\tau$ is the measurement time, $\lambda_d$ the dark count rate and $\lambda_s$ the signal rate. Contrary to Ref.~\cite{Mitridate:2020kly} in the following we assume that we are in the background dominated limit (long measurement time) and hence $\lambda_s<2\frac{\lambda_d}{\tau}$. The assumption of background domination is well justified with near future detector dark count rates in the case when we don't tune the resonances via the application of an external pressure. In case different axion masses are scanned by changing the external pressure on the material a shorter individual measurement time for each scan has to be assumed.

Using a background dominated single photon measurement the sensitivity is:
\begin{widetext}
\bea
	\g &>& \num{4.3}\times 10^{-12}\,{\rm GeV}^{-1} \left(\frac{0.01}{\eta}\right)^{\frac{1}{2}} \left(\frac{\SI{10}{\tesla}}{B_e}\right) \,  \left(\frac{100}{\beta}\right) \,  \left(\frac{(\SI{0.1}{\m})^2}{A}\right)^{\frac{1}{2}} \, \left(\frac{\lambda_\text{d}}{\SI{e-3}{\hertz}}\right)^{\frac{1}{4}}\, \left(\frac{3\,\text{yr}}{\tau}\right)^{\frac{1}{4}}\left(\frac{0.3\,{\frac{{\rm GeV}}{{\rm cm}^3}}}{\rho_{\rm DM}}\right)^{\frac{1}{2}} \, \left(\frac{m_a}{0.01\,{\rm eV}}\right)^{\frac{3}{2}},\nonumber\\
	\label{eq:sensitivity_backgroundDominated}
\eea
\end{widetext}
where $\eta$ is the photon counting efficiency, $\rho_{\rm DM}$ the dark matter density and $m_a$ is the axion mass. $A$ is the surface area of the material which we take to be $(0.1\,{\rm m})^2$.

Asides from any manufacturing constraints, there will be diminishing returns once the size of the system is larger than can be focused onto a single detector. This is because the transverse velocity of the axion leads to a small deviation from perfectly perpendicular emitted rays on the order of $\theta=10^{-3}$~\cite{Jaeckel:2015kea,Millar:2017eoc,Knirck:2019eug}. However, entendue (a combination of area and angular spread of light) is conserved. In other words, if the detector is in some refractive index $n_d$ and the area of the detector is given by $A'$, with light incident at some angle $\theta'$ then
\begin{equation}
A\sin^2\theta=n_d^2A'\sin^2\theta'\,,
\end{equation} 
meaning that $A\lesssim 10^6A'$ for a system in vacuum. The detectors we are considering have active areas of ${\cal O}(0.01-1)\,{\rm mm}^2$~\cite{BREAD:2021tpx}, leading to our choice of $A$. If one wishes to deviate beyond this limit, either materials with high refractive indices must be employed, or one will require multiple detectors, leading to a reduced signal to noise scaling. 

By optimizing the emitted electromagnetic field on resonance we can derive an optimal thickness that depends on the loss parameters:
\begin{widetext}
\bea
d_{\rm opt}=\frac{2}{\wLO}\left(\frac{ \Delta_j}{\sqrt{\ei}}\right)^{\frac{2}{3}}\left(\frac{1}{\frac{ \gr}{\wLO} +\frac{\gp \wLO\ei}{\wpl^2}}\right)^{\frac{1}{3}}= 0.2\,{\rm mm}\, \left(\frac{0.01\,{\rm eV}}{\wLO}\right)^{\frac{2}{3}}\,\left(\frac{\Delta_j}{\pi}\right)^{\frac{2}{3}}\,\left(\frac{2}{\sqrt{\ei}}\right)^{\frac{2}{3}}\,\left(\frac{10^{-4}\,{\rm eV}}{\gp}\right)^{\frac{1}{3}},
\label{eq:DA_AQ_P_optimalThickness}
\eea
\end{widetext}
where again neglecting photon losses we have set $\wLO=\wpl/\sqrt{\ei}$ and $\gr=\gp$ in the last equation to get a feeling for the order of magnitude of the optimal thickness. For higher frequencies $\wLO$ the optimal thickness is reduced even further. For smaller losses the optimal thickness is increased.

We can use Eq.~\eqref{eq:DA_AQ_P_mixing_LayerBoost_res_simp} with the optimal thickness from Eq.~\eqref{eq:DA_AQ_P_optimalThickness} to obtain a value for the boost factor on resonance. Plugging everything into the sensitivity estimate in Eq.~\eqref{eq:sensitivity_backgroundDominated} yields:
\begin{widetext}
\bea
	\g > \num{4.12}\times 10^{-11}\,{\rm GeV}^{-1}\, \left(\frac{0.01}{\eta}\right)^{\frac{1}{2}} \left(\frac{\SI{10}{\tesla}}{B_e}\right)    \left(\frac{(\SI{0.1}{\m})^2}{A}\right)^{\frac{1}{2}} \, \left(\frac{\lambda_d}{\SI{e-3}{\hertz}}\right)^{\frac{1}{4}}\, \left(\frac{3\,\text{yr}}{\tau}\right)^{\frac{1}{4}}\, , \nonumber \\ \times\left(\frac{0.3\frac{\rm GeV}{{\rm cm}^3}}{\rho_{\rm DM}}\right)^{\frac{1}{2}} \, \left(\frac{0.01{\rm eV}}{m_a}\right)^{\frac{1}{2}}\,\,\left(\frac{\Delta_j}{\pi}\right)^2\left(\frac{\SI{2}{\milli\metre}}{d}\right)^2\times \Sigma \, ,
	\label{eq:sensitivity_backgroundDominated_explicit}
\eea
\end{widetext}
where we have defined $\Sigma\equiv1+2\left(\frac{d}{d_\text{opt}}\right)^3$.

In order to make projections, we first performed a fit around each resonance \eqref{eq:dispersion_relation_simple} in which (as discussed in sec.~\ref{sec:phonon_Losses}) photon losses are neglected ($\Gamma_\rho =0$).   We took data from Ref.~\cite{Knapen:2021bwg} (which is collated in \cite{DarkElfGitHub}). The associated experimental references and measurement temperatures are displayed in table \ref{tab:losses}. Projections for the sensitivity to the axion-photon coupling are shown in Fig.~\ref{fig:Sensitivity} (note to avoid cluttering the plot, we only display the material sensitivities around the resonances where they exceed the dish). Integration time was set at $\tau = 1$ month for each resonance with $B_0 = 14 T$ and $A = (0.1 {\rm m})^2$. In all cases we also chose the optimal thickness of the material according to Eq.~\eqref{eq:DA_AQ_P_optimalThickness}, giving boost factors $\beta={\cal O}(1-100)$ depending on the assumed material properties. 

For the QCD model band, we show the 68$\%$ CL (orange) for the values of $C_{a\gamma \gamma}$ where $g_{a \gamma \gamma} = \alpha/(2\pi f_a) C_{a \gamma \gamma}$. The KSVZ \cite{Kim:1979if,Shifman:1979if} and DFSZ \cite{Dine:1981rt,Zhitnitsky:1980tq} QCD axion models are shown in yellow bands. 

The left and right regions of the plot with a discontinuous break correspond to QCDet \cite{Echternach:2018,Echternach2021} and SNSPD \cite{Hochberg:2019cyy,Verma:2020gso} single photon counters considered in \cite{BREAD:2021tpx}.  These operate in the frequency ranges $2-125$ meV and $124- 830$ meV, respectively.  Their respective dark count rates are $\lambda_d = 4$ and $\lambda_d = 10^{-4}$. 
 
Solid lines indicate pessimistic projections from material measurements done at temperatures  well above those encountered in an axion experiment (see table \ref{tab:losses}). Meanwhile, dashed lines display sensitivities achievable at low temperatures when only phonon decays dominate, leading to lower phonon losses. In that case, in the fitting function we replace $\Gamma \rightarrow \Gamma_{\rm dec, imp}$ with values given by those in Table~\ref{tab:losses} in accordance with the discussion in Sec.~\ref{sec:phonon_Losses}. The dashed lines also correspond to a  less-conservative dark count rate of 100 times lower than those quoted above. 

Finally, we note that similarly to many axion experiments, our setup also would be sensitive to other dark matter candidates. The most notable of these are hidden photons (also known as dark photons or paraphotons)~\cite{Arias:2012az}. Hidden photons are a new U(1) massive gauge boson with a small kinetic mixing with the visible photon~\cite{Fayet:1980rr,Okun:1982xi,Georgi:1983sy,Holdom:1985ag}. While some care is needed in analyzing the data, one could use the methods of Ref.~\cite{Caputo:2021eaa} to easily limit hidden photons. Further, for such a purpose no magnetic field is required, reducing one of the main cost drivers of an experiment.

\section{Conclusion}\label{sec:conclusion}

We have revisited the possibility, first suggested by Ref.~\cite{Mitridate:2020kly}, to detect axions via their interaction with phonon-polaritons in the materials GaAs, Al$_2$O$_3$, and SiO$_2$. We considered the conversion of phonon-polaritons to free photons at the material boundary, demonstrating that aixon DM should cause phonon-polariton materials to emit THz radiation in the presence of a strong magnetic field. We computed the signal power for disks of material of radius $\mathcal{O}(10\text{ cm})$ and thickness $\mathcal{O}(0.01\text{ mm})$, demonstrating signal enhancement compared to a magnetized mirror at the phonon-polariton resonance. We used measured and simulated dielectric functions of the phonon-polariton available in the literature to fit the loss parameters in the model. Measured losses in the 10-100 K range are relatively large, and lead to marginal enhancement compared to a simple mirror. We argued, however, that losses should be expected to decrease at sub-K temperatures available in axion DM searches with dilution refrigerators, and will saturate at a level $\Gamma/\omega\approx 5\times 10^{-4}$, set by the lattice spacing and crystal grain size. With such reduced losses, and with next generation THz detectors, phonon-polariton materials offer hope of axion DM searches exceeding astrophysical limits, and in some cases probing the QCD model band, for axion masses $m_a\in [30,160]\text{ meV}$. The small size of the samples (fixed by the axion de Broglie wavelength) may allow one to use $N$ samples in parallel, and thus further increase the signal.

Phonon-polariton materials are limited in their application to axion DM searches due to the difficulty of tunability, and depending on manufacturing requirements may have a smaller effective area than  dish antennas. However, if a broadband search for axions were to be successful in some range covering the phonon resonance, then phonon-polariton materials could be used to make more precise measurements and help to determine the axion mass, or as other independent confirmation of a signal, and would strongly motivate searches for resonant materials in the covered frequency range.

Further exploration of this method of axion DM detection requires precision measurement of the THz transmission spectra of the materials at very low temperature, in order to determine if the losses indeed saturate at the expected level. This can be performed by THz time-domain spectroscopy (TDS) in the range 3 to 16 THz. Fitting TDS measurements allows to determine the loss parameters and resonance frequencies of the materials in the relevant frequency range, and directly probes the channel of axion DM interaction via the source electric field. TDS measurements will also allow us to investigate the possibility of tuning the phonon-polariton resonance frequency by pressure or strain. Even a modest 10\% tuning range could greatly increase the utility of this method as a DM axion search. 

If tuning is impossible, or minimal, then extending this method further requires identifying more low-loss phonon-polariton materials in the THz range. Recently, Ref.~\cite{2021NatCo..12.1204L} has proposed new methods to discover phononic materials, and Ref.~\cite{2022arXiv220606248K} considered topological magnon materials. Similar efforts for phonon-polaritons may prove useful.

The viability of this method depends on the quality of THz detectors. We considered the same two benchmark THz SPDs as the BREAD proposal~\cite{BREAD:2021tpx}, namely QCDet \cite{Echternach:2018,Echternach2021} and SNSPD \cite{Hochberg:2019cyy,Verma:2020gso}. At their current dark count rates, neither is able to surpass the astrophysical limits on the axion photon coupling. Like BREAD, we also considered that next generation versions of these detectors may improve the dark count by up to two orders of magnitude, which allows us to surpass the astrophysical limits. Another relevant technology is the THz spectrometer of Ref.~\cite{Dona:2021iwt}, also being developed for axion DM searches.

We have revisited the approach to detect axion DM with phonon-polaritons. Contrary to previous approaches we have worked with the macroscopic axion-Maxwell equations and we have shown that low loss materials are necessary to reach the QCD axion band with near future detectors. This formalism is computationally simple and easily correlated with experimental measurements of the dielectric function, in addition to being able to handle finite boundary conditions. We computed the reach of three benchmark materials, however any materials with known dielectric functions and low losses may be useful for axion detection. Therefore in the future relatively low scale and low cost axion experiments could be carried out for specific axion masses by using phonon-polariton materials for which the dielectric functions have been measured.

\section*{Acknowledgments}

We acknowledge useful conversations with Christina Gao, Yonatan Kahn, Fabian R. Lux, Tanner Trickle, Kevin (Zhengkang) Zhang. DJEM is supported by an Ernest Rutherford Fellowship from the Science and Technology Facilities Council (UK). The work of JSE is supported in part by DOE grant DE-SC0015655. JSE would like to express special thanks to the Mainz Institute for Theoretical Physics (MITP) of the Cluster of Excellence PRISMA+ (Project ID 39083149), for its hospitality and support. JIM is supported by an FSR Incoming Postdoctoral Fellowship. AJM is supported by the European Research Council under Grant No. 742104 and by the Swedish Research Council (VR) under Dnr 2019-02337 “Detecting Axion Dark Matter In The Sky And In The Lab (AxionDM)”. Fermilab is operated by Fermi Research Alliance, LLC under Contract No. DE-AC02-07CH11359 with the United States Department of Energy.

\bibliography{bibliography}
\end{document}